\documentclass[aps,prl,superscriptaddress,twocolumn,showpacs,amsmath,amssymb]{revtex4}

\usepackage{color}
\usepackage{graphicx}
\usepackage{bm}
\definecolor{Blue}{rgb}{0.00, 0.00, 1.00}
\definecolor{Red}{rgb}{1.00, 0.00, 0.00}

\begin{document}

\title{Origin of photoresponse in black phosphorus photo-transistors}

\author{Tony Low$^{\dagger}$}
\email{tonyaslow@gmail.com}
\affiliation{IBM T.J. Watson Research Center, 1101 Kitchawan Rd., Yorktown Heights, NY 10598, USA}
\affiliation{Department of Physics and Electrical Engineering, Columbia University, New York, NY 10027, USA}
\author{ Michael Engel$^{\dagger}$ }
\affiliation{IBM T.J. Watson Research Center, 1101 Kitchawan Rd., Yorktown Heights, NY 10598, USA}
\affiliation{IBM Research-Brazil, Rio de Janeiro, RJ 22290-240, Brazil}
\author{ Mathias Steiner }
\affiliation{IBM Research-Brazil, Rio de Janeiro, RJ 22290-240, Brazil}
\author{ Phaedon Avouris}
\affiliation{IBM T.J. Watson Research Center, 1101 Kitchawan Rd., Yorktown Heights, NY 10598, USA}

%
\date{\today}

\begin{abstract}
We study the origin of photocurrent generated in doped multilayer BP photo-transistors, and find that it is dominated by thermally driven thermoelectric and bolometric processes. The experimentally observed photocurrent polarities are consistent with photo-thermal processes. The photo-thermoelectric current can be generated up to a $\mu$m away from the contacts, indicating a long thermal decay length. With an applied source-drain bias, a photo-bolometric current is generated across the whole device, overwhelming the photo-thermoelectric contribution at a moderate bias. The photo-responsivity in the multilayer BP device is two orders of magnitude larger than that observed in graphene.\\
$\dagger$ \emph{These authors contributed equally}
\end{abstract}
\maketitle

\emph{Introduction---} Like graphene, black phosphorus (BP) is also a layered material, except that each layer forms a puckered surface due to its $sp^3$ hybridization. The electrical, optical and structural properties of single crystalline and polycrystalline BP had been studied in the past\cite{Keyes53,Warschauer63,Jamieson63,morita86review,chang86}. Recently, interests in BP has re-emerged\cite{Li14BP,Liu14BP,xia14bp,Koenig14,Castellanos14,rodin14,Rudenko14} in its multilayer thin film form, obtained by simple mechanical exfoliation\cite{novoselov2d05}. In its bulk form, BP is a semiconductor with a direct band gap of about $0.3\,$eV. In addition, the optical spectra of multilayer BP also vary with thickness, doping, and light polarization across mid- to near-infrared frequencies\cite{low14bpcond}. Bulk BP exhibits excellent electrical properties compared with other layered semiconductors, with measured Hall mobilities in $n$ and $p-$type samples approaching $10^5\,$cm$^2$/Vs\cite{morita86review}. In addition, recent electrical data on multilayers BP thin films showed encouraging results with mobility of $1000\,$cm$^2$/Vs, making it  an attractive alternative to narrow gap compound semiconductors for optoelectronics across mid- to near-infrared frequencies.

In this work, we study the origin of photoresponse in BP thin film photo-transistors. Photoexcitation at energies far above the energy gap produces electrons (and holes) with large excess energy. Conversion of the excess energy of these photoexcited carriers into electrical current before they dissipate into the thermal sinks represents one of the key challenges to efficient optoelectronic devices. Energy relaxation of the photoexcited carriers predominantly occurs via different inelastic scattering channels such as, intrinsic optical and acoustic phonons\cite{Bistritzer09}, or  remote surface polar phonon modes of the substrate\cite{low12cool}. These processes can produce  elevated local electronic and phononic temperatures which subsequently drives a thermal-current, i.e. by thermoelectric\cite{lamme11te,gabor11hot} and bolometric\cite{freitag12photo,freitag13photo} processes.

In the photo-thermoelectric process, the photoresponsivity $R_ {TE}\equiv I_{TE}/P$, which is defined as the generated photocurrent per unit incident laser power, depends on various materials properties. Here, $R_{TE}\sim \sigma S/(\kappa_e+\kappa_{ph})$, where $\sigma$ is the electrical conductivity, $S$ is the Seebeck coefficient, and $\kappa_e$ ($\kappa_{ph}$) are the electronic (phononic) thermal conductivities. In black phosphorus, its high electrical conductivity coincides with a low in-plane lattice thermal conductivity of $12.1\,$Wm$^{-1}$K$^{-1}$ at room temperature\cite{slack65bp}. The latter is attributed to the large anharmonicity of the in-plane phonon modes, and the low sound velocity of the acoustic modes\cite{Morita89aniso}. Similar attributes are also found in anisotropic layered SnSe crystals, a good material candidate for thermoelectrics\cite{zhao14te}. On the other hand, graphene is a poor thermoelectric material because of its high lattice thermal conductivity i.e. $>2000\,$Wm$^{-1}$K$^{-1}$\cite{balandin11}. The Seebeck coefficient, $S$, of BP is also estimated to be larger than graphene\cite{Lv14te,fei14bpte,qin14tebp}.


In photo-bolometric process, local heating by the laser produces a differential change in resistance, which can be detected in a typical photo-conductivity setup\cite{freitag12photo,freitag13photo}. The photoresponsivity varies as $R_{B}\sim \beta/(\kappa_e+\kappa_{ph})$, where the bolometric coefficient quantifies the sensitivity of the electrical conductivity with temperature i.e. $\beta\equiv d\sigma/dT$. At room temperature, its carrier mobility is dominated by acoustic phonon scattering, which has a $T^{-3/2}$ temperature dependence for bulk\cite{morita86review}. In the 2D limit, one would expect the acoustic phonon limited mobility to have $T^{-1}$ dependence like graphene\cite{chen08ph}, when $T$ is larger than its Bloch Grun$\ddot{e}$isen temperature. However, experiments with BP multilayer have found an anomalous temperature dependence of $T^{-1/2}$ instead, reminescent to the 1D case, which can most probably attributed to the highly anisotropic bandstructure of BP\cite{Li14BP,Liu14BP,xia14bp}. Simple estimates based on $R_B \propto \gamma_s^2/D_{ac}^2$, where $D_{ac}$ is the deformation potential and $\gamma_s$ is the Grun$\ddot{e}$isen parameter, would also suggests a larger $R_B$ in BP than graphene.

In this letter, we show that the photoresponse in BP photo-transistor is dominated by thermally driven processes. We begin with a brief description of our device electrical characteristics, from which its Seebeck coefficient and relevant transport coefficients can be extracted. Electronic and phononic temperatures of the device are modeled via coupled differential heat transport equations. Thermally driven thermoelectric and bolometric currents, and their polarities, can then be estimated, and compared against the measured photocurrent as a function of applied bias and laser intensity. The experimentally observed photocurrents polarities and dependencies are consistent with our picture of thermally driven thermoelectric and bolometric processes, not with the photovoltaic effect.

\begin{figure}[t]
\centering
\scalebox{0.5}[0.5]{\includegraphics*[viewport=115 60 700 515]{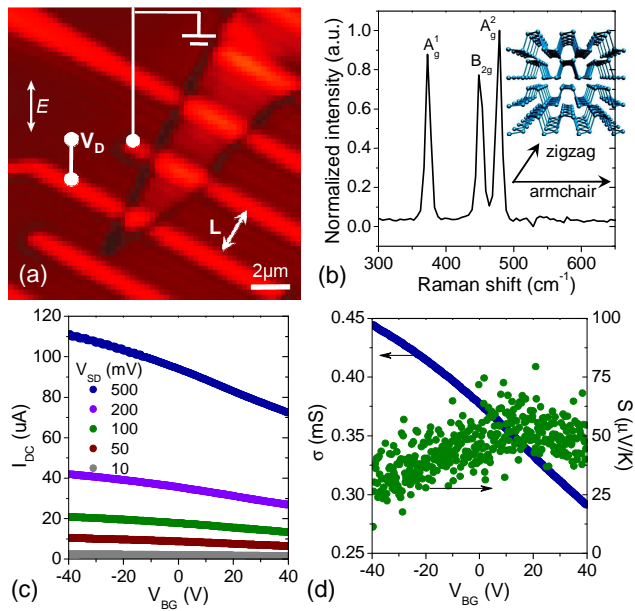}}
\caption{
\textbf{(a)} Laser reflection image ($15\,\mu$m $\times 15\,\mu$m) of the BP device. Source and drain terminals are indicated. The light polarization used for the photocurrent and Raman spectroscopy are indicated. \textbf{(b)} Raman spectra of BP showing the prominent representative normal modes of the $\Gamma$ point optical phonons. Three other Raman-active modes are not observed because of selection rules for the scattering configuration\cite{Sugai85}.  \textbf{(c)} Electrical transfer characteristic of the device measured at different source drain voltages $V_{SD}$, averaged over the positive and negative back gate voltages $V_{BG}$ sweep.
\textbf{(c)} Electrical conductivity $\sigma$ and the Seebeck coefficient $S$ extracted from the measured transfer characteristics, see text for details. 
}
\label{figure1}
\end{figure}

\emph{Device characteristics---} The layered structure of BP allows for mechanical exfoliation into multilayer structures on Si/SiO$_2$ substrates. Fig.\,\ref{figure1}a shows the laser reflection image of the BP device, which has a channel length of $L\approx 2\,\mu$m and an averaged width of $W\approx 1\,\mu$m. The black phosphorus multilayer has a thickness of $\approx 100\,$nm, based on atomic force microscope measurement. Fig.\,\ref{figure1}b shows the Raman spectrum of the device, from which we determined that our device channel is oriented $15^o$ with respect to the armchair direction of the crystal axes (see inset)\cite{xia14bp}. The device is contacted by Pd  leads. Thicker BP films offer the benefit of higher light absorption and carrier mobilities, both crucial attributes for photodetection. Fig.\,\ref{figure1}c shows the electrical transfer characteristic of the device measured at different source drain voltages $V_{SD}$, averaged over the positive and negative back gate voltages $V_{BG}$ sweep. The current modulation with $V_{BG}$ is rather moderate as expected, since the BP thickness is larger than the out-of-plane screening length of order $\sim 10\,$nm\cite{low14bpplas}. The current sourced exhibits a linear dependence with the applied $V_{SD}$, allowing us to extract the electrical conductivity $\sigma$ as shown in Fig.\,\ref{figure1}c. The $V_{BG}$ dependence of $\sigma$ indicates that our BP device is p-doped.

The Seebeck coefficient is related to the electrical conductivity $\sigma$ via the Mott formula\cite{ashcroft76}, and can be obtained from the experimental $\sigma$,
\begin{eqnarray}
S &=& -\frac{\pi^2 k_B^2 T}{3C_{ox}}\left(\frac{1}{\sigma}\frac{d\sigma}{dV_g}\right)\frac{dn}{d\epsilon_f}
\end{eqnarray}
where $\epsilon_f$ is the Fermi energy, $n$ is the electron density and $C_{ox}$ is the back gate oxide capacitance ($300\,$nm SiO$_2$ dielectric). Quantities involving $\sigma$, as expressed within the bracket, can be obtained from experimentally measured $\sigma$. For BP multilayer thin films, its carrier density can be computed within the effective mass framework as described in Ref.\,\cite{low14bpcond}, from which $dn/d\epsilon_f$ can be computed. 
Fig.\,\ref{figure1}d plots the extracted Seebeck coefficient across the applied $V_{BG}$. $S$ is positive since the device is p-doped, and the dependence with doping is consistent with the expected behavior, that the magnitude of $S$ decreases with doping. With the measured electrical conductivity of $\sigma\approx 0.4\,$mS and taking the hole mobility to be $\approx 1000\,$cm$^2$/Vs\cite{Li14BP,Liu14BP,xia14bp}, we arrive at a considerable hole carrier density of $2.5\times 10^{12}\,$cm$^{-2}$. The magnitude of $S\approx 60\,\mu$V/K is comparable to that observed in graphene\cite{zuev09}, but is smaller than that estimated in recent calculations\cite{qin14tebp}, which can probably be attributed to the presence of disorder in our samples.  

\begin{figure}[t]
\centering
\scalebox{0.5}[0.5]{\includegraphics*[viewport=160 50 700 565]{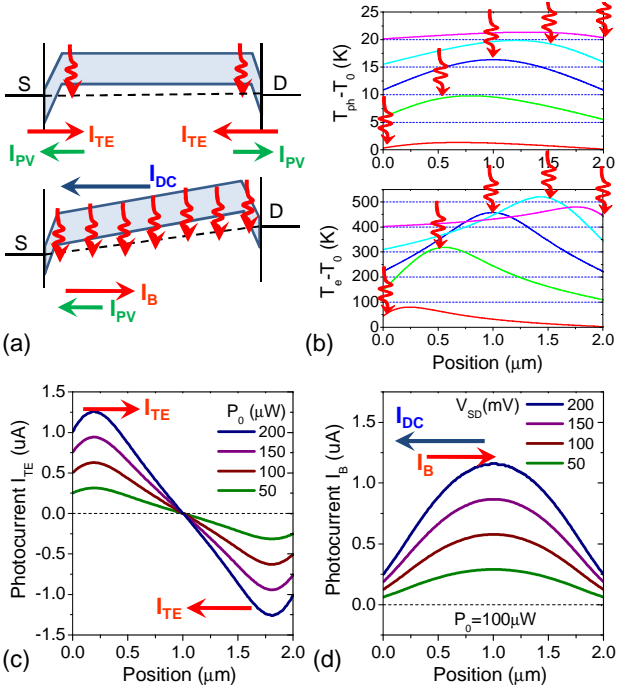}}
\caption{
\textbf{(a)} Energy band diagram of the device at zero (top) and finite (bottom) source-drain bias. The polarities of the various photocurrents i.e. thermoelectric, bolometric, photovoltaic are indicated. \textbf{(b)} Simulated spatial profiles of the elevated electronic and phonon temperatures (i.e. $T_e$ and $T_{ph}$) due to local laser induced heating as indicated. Ambient temperature $T_0$ is taken to be $300\,$K. \textbf{(c)} Simulated laser-scanned photo-thermoelectric currents at different incident power.  \textbf{(c)} Simulated laser-scanned photo-bolometric currents at different applied source-drain bias. 
}
\label{figure2}
\end{figure}

\emph{Modeling---} The polarities, and order of magnitude estimates of photo-thermoelectric and photo-bolometric effect are presented, which can be compared against the experiments in the next section.

Light absorbed by the BP device is modeled with a power density following a Gaussian profile as follows,
\begin{eqnarray}
P(x)=\frac{\alpha P_0}{a_0 L_s}\mbox{exp}\left[-\frac{(x-x_0)^2}{2a_0^2}\right]
\end{eqnarray}
where $a_0=L_s/2\times \sqrt{ 2\pi/2\mbox{log}2  }$ is the spread in terms of the laser spot size $L_s$, $P_0$ is the incident power and $\alpha$ is the absorption coefficient. For infrared frequencies, $\alpha\approx 0.005$ per nm\cite{low14bpcond} in BP along the armchair direction, which translates to $\alpha\approx 0.5$ for our device. 


Each thermal bath can be characterized by their respective temperatures. Electronic ($T_e$) and phononic ($T_{ph}$) temperatures of the device are modeled via a coupled differential heat transport equations as follows,
\begin{eqnarray}
\nonumber
-t\kappa_e \frac{\partial^2 T_e}{\partial x^2}+\gamma_{e-ph}(T_e-T_{ph})&=&P(x)\\
-t\kappa_{ph} \frac{\partial^2 T_{ph}}{\partial x^2}+\gamma_{0}(T_{ph}-T_0)&=&\gamma_{e-ph}(T_e-T_{ph})
\end{eqnarray}
where $T_0=300\,$K is the ambient temperature. The Si substrate and metallic contacts, which serve as heat sinks are at $T_0$. The electronic thermal conductivity, $\kappa_e$, can be determined from the Wiedemann Franz law from the measured $\sigma$, and was found to be $0.5\,$Wm$^{-1}$K$^{-1}$. The lattice thermal conductivity, $\kappa_{ph}$, was experimentally found to be $12.1\,$Wm$^{-1}$K$^{-1}$ for bulk polycrystalline samples at $300\,$K\cite{slack65bp}. We adopt this measured value for our calculation, but note that moderate anisotropy to within an order of magnitude, should be expected in crystalline samples. 

Besides the in-plane heat transport, there are also heat exchanges with the substrate. Heat flow into the underlying Si substrate is mediated by the SiO$_2$ dielectric of thickness $300\,$nm. In addition, the finite BP thickness implies also an out-of-plane thermal resistance. With an out-of-plane lattice thermal conductivity of $\approx 1\,$Wm$^{-1}$K$^{-1}$ in BP\cite{slack65bp}, and a  bulk thermal conductivity $0.5-1.4\,$Wm$^{-1}$K$^{-1}$ in SiO$_2$\cite{yamane02}, this translates to an effective out-of-plane thermal conductivity of order $\gamma_0\approx 1\,$MW/Km$^2$ for the BP-SiO$_2$ stack.

Typically, electron cooling rate at room temperature is dominated by inelastic scattering processes with acoustic phonons. In the temperature regime of interest, where Maxwell– Boltzmann statistics is approximately applicable, the electron energy-loss rate via acoustic phonons is known to be linear in temperature\cite{ridley91}. Hence, we can express this energy loss rate via a thermal conductivity,  $\gamma_{e-ph}$. Currently, there are no estimates for  $\gamma_{e-ph}$ in BP. Hence we tentatively assigned a value of $\gamma_{e-ph}\approx 0.1\,$MW/Km$^2$, similar to graphene\cite{low12cool}, which we later found to provide good agreement to the experimentally measured photocurrent.




The steady state current due to local heating in our device can be written as,
\begin{eqnarray}
\label{currentcom}
&&I= \sigma (V_d-V_s)+ \int_0^L \sigma S(x) \frac{dT_e(x)}{dx} dx \\
\nonumber
&&- \int_0^L e\mu n^{*}(x) \frac{dV(x)}{dx} dx + \int_0^L  \beta (T_{ph}(x)-T_0)\frac{dV(x)}{dx}dx 
\end{eqnarray}
where $n^{*}$ is the photo-excited carriers density, $V(x)$ is the applied electric potential, while $E(x)$ is the induced electric field in response to the photo-excitation that establishes current continuity\cite{song11hot}. The current components in Eq.\,\ref{currentcom} are the dark current ($I_{DC}$), photo-thermoelectric ($I_{TE}$), photovoltaic ($I_{PV}$), and bolometric ($I_B$) respectively.


\emph{Photo-thermoelectric---} Fig.\,\ref{figure2}a (top) illustrates the energy band diagram under zero applied bias. Our device has an electrical conductivity of $\sigma\approx 0.4\,$mS and Seebeck coefficient $S\approx 60\,\mu$V/K. The metallic contacts, on the other hand, have very poor thermoelectric sensitivities i.e. $S\approx 0\,\mu$V/K. The metal-BP junction therefore effectively acts as a thermocouple, which upon heating would produce a hole current flowing into the BP channel as depicted in Fig.\,\ref{figure2}a. 


Fig.\,\ref{figure2}b shows the computed temperatures profiles, $T_e$ and $T_{ph}$, due to local excitation by a continuous wave laser at different locations as indicated. For photo-thermoelectric effects, $T_e$ is the relevant temperature. Maximal heating of the metal-BP junction occurs when the laser is parked directly above, with $T_e-T_0\approx 50\,$K. We observed that the thermal energy flows for more than a $\mu$m, with $T_e-T_0\approx 0\,$K at the opposite metal-BP junction. Hence, when the laser is parked at the middle of the channel, heating of the metal-BP junction can also occurs, producing an elevated electronic temperature $T_e-T_0\approx 20\,$K. This long thermal length of $1\,\mu$m (determined by the electron-phonon coupling $\gamma_{e-ph}$) can produce a seemingly nonlocal photo-thermoelectric current $I_{TE}$. Fig.\,\ref{figure2}c shows the simulated $I_{TE}$ as the laser scans across the length of the device. Indeed, finite photoresponse persists even $\mu$m away from the photo-active metal-BP junction, and a zero response is obtained when the laser is at the middle of the symmetric channel due to canceling photocurrents from the two contacts. Maximal $I_{TE}$ occurs at finite distance from the junction due to tradeoffs between the dual functions of metal contact as a photoactive junction and a heat sink. $I_{TE}$ increases proportionally with increasing laser power $P_0$.

\emph{Photo-bolometric---} Fig.\,\ref{figure2}a (bottom) illustrates the energy band diagram for our BP device with finite applied bias. Photo-induced heating can modify the electrical transport cofficients, leading to an electrical conductivity that differs from that in the dark. The bolometric coefficient quantifies the sensitivity of the electrical conductivity with temperature i.e. $\beta\equiv d\sigma/dT$. Experiments in BP multilayers found a temperature dependence of $T^{-1/2}$ due to acoustic phonon scattering\cite{Li14BP,Liu14BP,xia14bp}. With a dark electrical conductivity of $\sigma\approx 0.4\,$mS, we arrived at $\beta\approx  -0.7\,\mu$SK$^{-1}$. Negative $\beta$ implies a negative photo-conductivity, typical in metallic or doped materials\cite{freitag12photo,frenzel14}. 

In the case of photo-bolometric effect, $T_{ph}$ is the relevant temperature. As shown in Fig.\,\ref{figure2}b, $T_{ph}$ exhibits a similar behavior as $T_{e}$, except an order of magnitude smaller i.e. $\sim 10\,$K. This is expected since the thermal resistance between the electron and phonon baths is an order of magnitude larger than the phonon and substrate i.e. $\gamma_{e-ph}^{-1}\ll \gamma_{0}^{-1}$. Fig.\,\ref{figure2}d shows the simulated bolometric current $I_{B}$ as the laser scanned across the length of the device. The $I_{B}$ flows in opposite direction to the $I_{DC}$, and has the largest magnitude in the middle of the channel, and increases linearly with $V_{DS}$.

\emph{Photo-photovoltaic---} Local electric field can drive the photo-excited $n^*$, producing a photocurrent. However, our BP device has substantial p-doping, and electron-electron scattering can significantly reduce $n^*$. Indeed, in graphene, one observed a crossover from  photovoltaic to bolometric as doping increases\cite{freitag12photo,frenzel14}. Based on the metal gate stacks we used (Ti/Pd/Au), and the substantial p-doping in BP, we expect a Schottky junction at the metal-BP interface as depicted in Fig.\,\ref{figure2}a. In both the zero and finite bias case, the photovoltaic current would bear opposite polarity to the thermoelectric and bolometric currents, allowing for direct experimental verification of the photocurrent origins via their polarities.

\begin{figure}[t]
\centering
\scalebox{0.5}[0.5]{\includegraphics*[viewport=125 95 700 560]{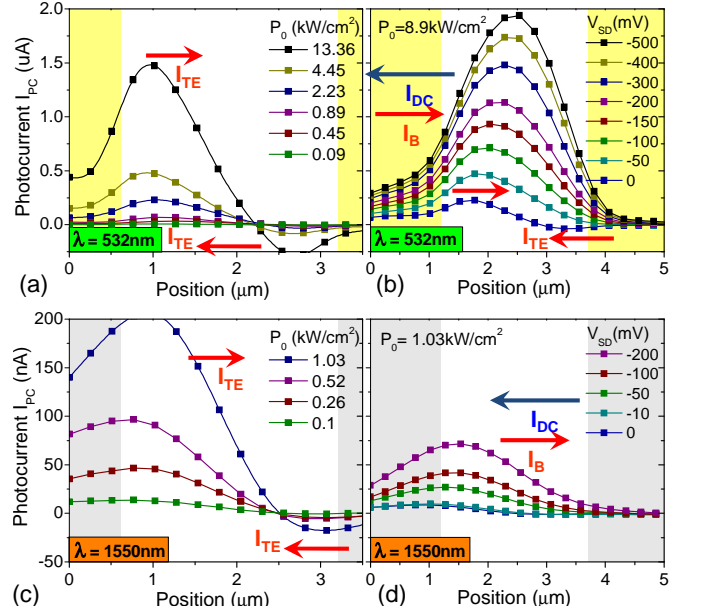}}
\caption{
Photocurrent cross-sections extracted from photocurrent maps taken along the direction of maximum amplitude measured as a function of optical power density. Photocurrent is laser excited at visible wavelength of $532\,$nm, at different \textbf{(a)} incident power and \textbf{(b)} applied source-drain bias.  Similar measurement are done for infrared wavelength of $1550\,$nm in  \textbf{(c-d)}.  Shaded areas indicate contact regions of the device.
 }
\label{figure3}
\end{figure}

\emph{Measured photoresponse---} A focused laser beam at visible wavelength of $532\,$nm is scanned across the channel of the device by a piezo-driven mirror to acquire the spatial photocurrent profiles. The photocurrent setup and microscopy is discussed elsewhere\cite{engel14bp}.  Fig.\,\ref{figure3}a and b plots the measured photocurrent spatial profiles as function of laser intensity and applied bias respectively. The current polarities follow from the device energy band diagram as shown in Fig.\,\ref{figure2}a, where current flowing from source to drain direction is assigned as positive. It is immediately apparent that the measured photocurrent polarities are consistent with the thermoelectric and bolometric processes, while the photovoltaic effect predicts the opposite polarity. In addition, the photo-thermoelectric current can be picked up a micron away from the photoactive contacts, indicative of the long thermal decay length, consistent with photo-thermoelectric effect discussed earlier. On the other hand, photovoltaic effect would only be observed where there is local electric fields i.e. at the contacts.

A laser power of $100\,\mu$W translates to a power density of $\approx 20\,$kW/cm$^2$. The maximal observed photo-thermoelectric current (average of the two junctions) is of the same order as calculated, but with strong asymmetry due to the device geometry. Normalizing it to the total incident power, the photo-responsivity is $\approx 20\,$mA/W. This larger responsivity, about two orders of magnitude larger than its graphene counterparts\cite{freitag13sus}, can be largely attributed to the larger $\alpha$ associated with the $100\,$nm BP film. Compensating for the larger $\alpha$, about $50\%$ in our BP device versus $2\%$ in monolayer graphene, one arrives at a `re-normalized' photo-responsivity an order larger still. 

With an applied source-drain bias, a photo-bolometric current is generated across the whole device, eventually overwhelming the photo-thermoelectric currents at moderate bias of $0.5\,$V. The measured photo-bolometric current is $\sim 2\times$ larger than that calculated. This would suggest a slightly larger electron-phonon coupling $\gamma_{e-ph}$, or bolometric coefficient $\beta$ than that assumed in our analysis. We note that the $\beta$ used is similar to that obtained in graphene\cite{freitag12photo}. We also observe a slight tilt in the maximal bolometric current towards the drain contacts, probably due to drain-voltage induced doping of the channel interior. Similarly, the observed photo-responsivity due to bolometric effect is about two orders of magnitude larger than that obtained in graphene\cite{freitag12photo}. We observed similar photocurrent behavior and responsivity at infrared wavelength of $1550\,$nm, as shown in Fig.\,\ref{figure3}c and d.

\emph{Conclusion---} In conclusion, we studied the origin of photocurrents in doped multilayer BP photo-transistors, and found that it is dominated by thermally driven thermoelectric and bolometric processes. The experimentally observed current polarities are consistent, while the photovoltaic currents would predict otherwise.  Multilayer BP offers an attractive alternative to narrow gap compound semiconductors for optoelectronics applications involving hyperspectral light detection covering both visible and infrared frequencies.

\emph{Note---} While preparing this manuscript, we became aware of recent works that discusses the origin of photocurrent in BP photo-transistors, due to photothermoelectric\cite{hong14te} and photovoltaic\cite{Buscema14pv,Buscema14}.


\end{document}